# Designing inorganic semiconductors with cold-rolling processability

Xu-Dong Wang[1#], Jieling Tan[1#], Jian Ouyang[1], Hangming Zhang[1], Jiang-Jing Wang[1], Yuecun Wang[2], Volker L. Deringer[3], Jian Zhou[1], Wei Zhang[1*], En Ma[1*]

[1]*Center for Alloy Innovation and Design (CAID), State Key Laboratory for Mechanical Behavior of Materials, Xi'an Jiaotong University, Xi'an 710049, China*
[2]*Center for Advancing Materials Performance from the Nanoscale (CAMP-Nano) & Hysitron Applied Research Center in China (HARCC), State Key Laboratory for Mechanical Behavior of Materials, Xi'an Jiaotong University, Xi'an 710049, China*
[3]*Department of Chemistry, Inorganic Chemistry Laboratory, University of Oxford, Oxford, OX1 3QR, UK*

[#]These authors contributed equally to this work.

*Emails: wzhang0@mail.xjtu.edu.cn , maen@xjtu.edu.cn

**Abstract**

While metals can be readily processed and reshaped by cold rolling, most bulk inorganic semiconductors are brittle materials that tend to fracture when plastically deformed. Manufacturing thin sheets and foils of inorganic semiconductors is therefore a bottleneck problem, severely restricting their use in flexible electronics applications. It was recently reported that a few single-crystalline two-dimensional van der Waals (vdW) semiconductors, such as InSe, are deformable under compressive stress. Here we demonstrate that intralayer fracture toughness can be tailored via compositional design to make inorganic semiconductors processable by cold rolling. We report systematic *ab initio* calculations covering a range of van der Waals semiconductors homologous to InSe, leading to material-property maps that forecast trends in both the susceptibility to interlayer slip and the intralayer fracture toughness against cracking. GaSe has been predicted, and experimentally confirmed, to be practically amenable to being rolled to large (three quarters) thickness reduction and length extension by a factor of three. Our findings open a new realm of possibility for alloy selection and design towards processing-friendly group-III chalcogenides for flexible electronic and thermoelectric applications.

Inorganic semiconductors are rapidly expanding their utility in flexible electronic applications, such as wearable devices, bendable displays, e-papers and others.[1-7] The widespread exploitation of these exciting opportunities, however, often requires large-area thin sheets of the respective material. So far, flexible electronic devices have mainly been based on organic semiconductors, or on inorganic semiconductor films vapor-deposited on flexible substrates.[8-13] While 2D pieces can be exfoliated from 3D single crystals, the resulting flakes are difficult to reproduce, highly fragile and very small in size. To overcome these three roadblocks, there is a pressing need for the design of inorganic semiconductors that are amenable to plastic deformation. In particular, it is highly desirable that they can be thinned and reshaped via a conventional thermo-mechanical processing route, such as the large-scale rolling operations that are commonly used for producing metal sheets. This has remained a daunting challenge so far for inorganic semiconductors, imposing severe limitations on their forming capability and applications. [14]

Recently, several inorganic semiconductors, including α-$Ag_2S$, β-InSe and $CrI_3$, have been reported to possess noteworthy malleability at ambient conditions.[15-18] The atoms in these materials are packed in layered hexagonal structures, each atomic layer being either flat or puckered, with strong intralayer covalent bonding, and weak (van der Waals like) but non-negligible interlayer interactions. This bonding hierarchy allows interlayer slippage while maintaining the integrity of the bulk phase, therefore imparting some degree of plastic deformation under applied stress.[19-20] This has led to proposals of their use in devices: α-$Ag_2S$ in flexible thermoelectric devices and temperature sensors,[21-25] InSe in field-effect transistors, photodetectors, and thermoelectric generators,[26-39] and $CrI_3$ possibly for high-density storage as a two-dimensional magnetic material.[40-41] However, it remains unclear whether any of these "deformable inorganic semiconductors" can survive rolling-induced shaping[14] without catastrophic fracture, which will be examined in **Figure 1** below.

As shown in Figure 1a, the hexagonal β-InSe takes a layered structure connected by van der Waals (vdW) interactions ($P6_3/mmc$ space group). In each atomic block, In and Se atoms form two bond-sharing tetrahedral motifs via In−In bonds. Following Ref. [16], we prepared β-InSe samples via the Bridgeman method (see Methods section). X-ray diffraction (XRD) and transmission electron microscopy (TEM) experiments confirm that the as-grown sample was in the β-phase (Figure S1). The InSe sample, initially 0.262 mm in thickness and ~5 mm in length, was subjected to continuous rolling (steel rollers) at room temperature, a typical mechanical processing route for ductile metals. The sample was gradually thinned down via 10 repeated rolling passes, with a thickness reduction of 0.02 mm each pass, eventually reaching a thickness of 0.065 mm together with an extension in length to 15 mm (Figure 1b). Scanning electron microscopy (SEM) images showing the morphology of the sample/surface before and after rolling are displayed in Figure 1b. Numerous small cracks were observed on the surface after rolling. Although surface polishing can help to strip off (some of) the fractured flakes, this rolling outcome is clearly unacceptable, both in terms of the mechanical integrity of the processed foil and the functional (electrical) performance of the post-rolling sample that is now flaw-ridden. In other words, the deformability claimed for these inorganic semiconductors in bending of thin foils or micrometer-sized-pillar tests[16] appears to be still far from practical for forming operation of bulk materials.

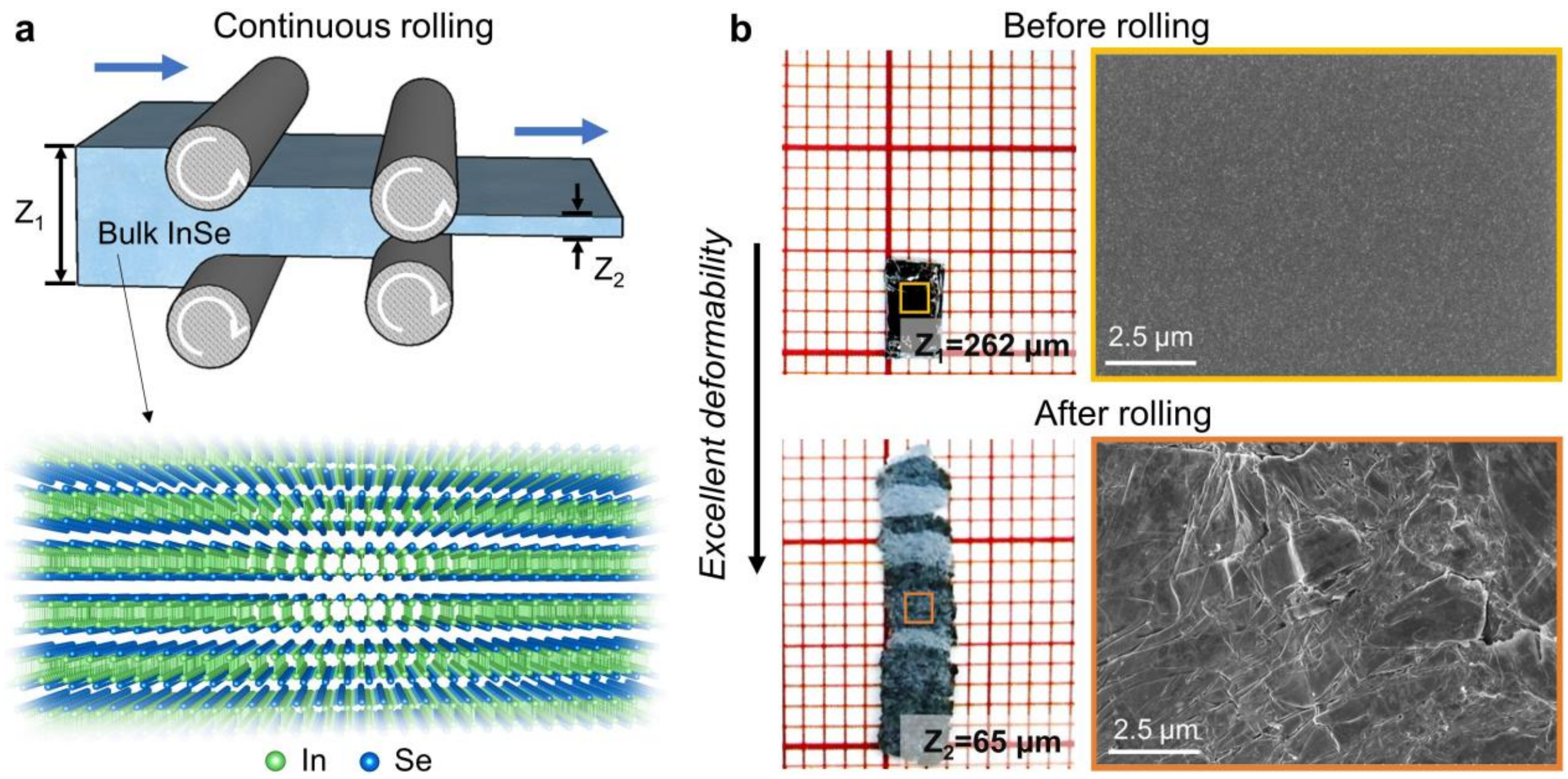

**Figure 1. Rolling pressing of bulk crystalline InSe. a** Schematic diagram showing the continuous rolling process. The inset shows the layered atomic structure of the InSe crystal. **b** The InSe sample (left panel) and the corresponding scanning electron microscopy image of its surface morphology (right panel) before and after rolling, respectively.

To gain insight into the key material properties affecting the rolling deformation/fracture, we carried out high-resolution TEM experiments to assess the fracture details. We applied mechanical exfoliation to induce fracture in the TEM specimen.[42-43] As sketched in **Figure 2**a, we exfoliated the sample with Scotch tape, and transferred the flakes to the TEM grid for characterization. Figure 2b shows the TEM image and the corresponding fast Fourier transform pattern recorded for a typical flake. Edges along the (100) and ($\bar{2}10$) planes were observed, corresponding to the zigzag and armchair edges (Figure 2c), respectively. More than 10 flakes were analyzed for statistics; no strong preference of zigzag edges over armchair edges or vice versa was identified. This can be understood from the comparable surface formation energy of the two types of crack edges, i.e. 0.058 eV/Å$^2$ for zigzag edge and 0.052 eV/Å$^2$ for armchair edge, as calculated using density functional theory (DFT; see Methods section).

Next, we produced the ideal biaxial tensile stress–strain curve of β-InSe using DFT. As shown in Figure 2d, the calculated ultimate strength and strain are ~9 GPa and ~12%, respectively, in good agreement with experimental data (8.68 GPa and 8.57%).[44] A nanoindentation test data point is also included for comparison. Based on the ideal stress–strain curve, we predict the fracture toughness *K* according to the Griffith–Irwin relation,[45]

$$K = \left(\frac{GE}{1-\upsilon^2}\right)^{1/2},$$

where *E* is the Young's modulus, $\upsilon$ is the Poisson's ratio, and *G* is the energy release rate, which can be calculated as $G = \int_{l_0}^{l_{\mathrm{limit}}} \sigma \, dl$, where $\sigma$ is the tensile stress, and $l_0$ and $l_{\mathrm{limit}}$ are the initial and limit length of the sample, respectively.[46-48] This approach to simulate tensile behavior has been used before for many single-crystalline thermoelectric materials, yielding good agreement with experimental values.[48] Our DFT calculation shows a much larger fracture toughness of β-InSe, viz.

0.15 MPa·m$^{0.5}$, as compared to the typical chalcogenide thermoelectric materials, such as $Bi_2Te_3$ and SnSe, for both of which this value is smaller than 0.03 MPa·m$^{0.5}$.[48]

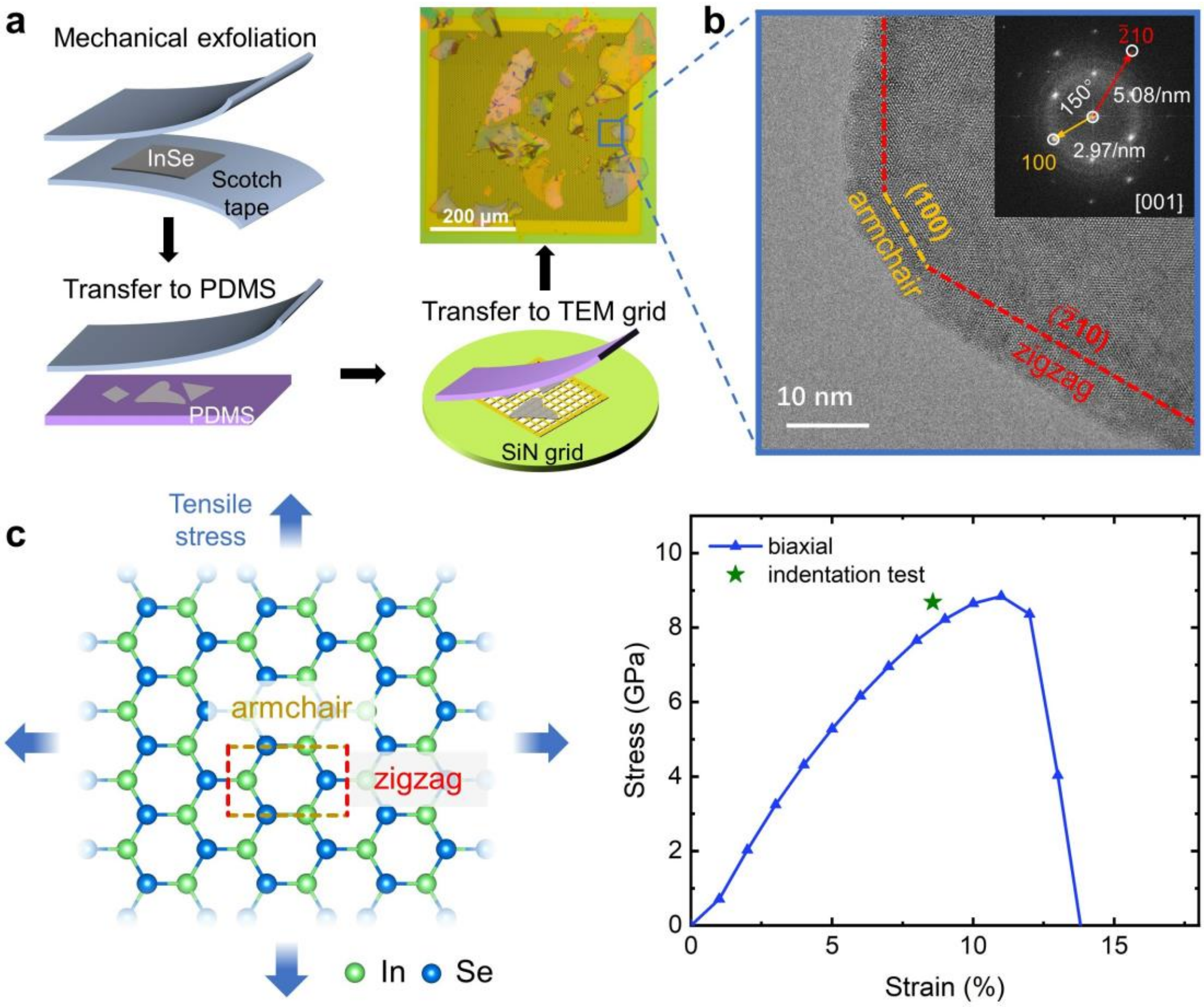

**Figure 2. Crack edges and stress-strain curve for InSe. a** Schematic diagram showing the approach to prepare crack edges of a crystalline InSe sample. PDMS is short for polydimethylsiloxane. **b** High-resolution transmission electron microscopy characterization for one typical crack edge. The inset shows the fast Fourier transform pattern. The ($\bar{2}10$) and (100) planes correspond to the zigzag and armchair type of edges, respectively. **c** Calculated stress–strain curve under biaxial tensile stress (schematically illustrated in the left panel). The experimental ultimate strain and tensile stress data point (green) from a nanoindentation test is from Ref. [44].

We note that in the origami[16] or rolling processes of β-InSe, the stress concentration can be relieved, and the permanent shape change is accommodated, by interlayer slipping. That is, the stress drives bond-switching slip between adjacent atomic layers rather than forcing excessive in-plane strain. This interlayer slip provides a desired mode to provide large extensibility, before the onset of runaway damage. In other words, although the fracture toughness of β-InSe is smaller than that of some familiar semiconductors (e.g. 0.8 MPa·m$^{0.5}$ for silicon)[48], no catastrophic damage has been initiated. That is, large plastic strain is achievable upon roller-constrained deformation via repeated interlayer slip.

The question then becomes how to suppress intra-layer cracking while sustaining the interlayer slip. Our goal is to allow sufficiently effective in-plane energy dissipation to delay notch formation, while sustaining interplane (weak) bond switching to mediate sliding. This obviously requires a judicious choice in balancing and optimizing the relative strength of interlayer vs. intralayer chemical bonding in the layered hexagonal structure (**Figure 3**). We here advocate for a rational design of the chemical interactions involved by tuning the chemical species in the semiconductor compound, moving up or down within a group of the Periodic Table. To this end, we performed DFT calculations of a series of MX alloys homologous to InSe, with the group-13 element being M = Al, Ga or In, and the group-16 (chalcogen) element being X = S, Se or Te (Figure 3). DFT-relaxed lattice parameters and typical interatomic distances of all nine compounds are given in **Table 1**. All these MX alloys are dynamically stable, as reflected by the absence of imaginary frequencies in the DFT-calculated phonon dispersions: results for GaSe and InS (obtained by substituting the cation or anion in InSe, respectively) are shown in Figure 3b; phonon dispersions for the other alloys studied are shown in Figure S2. All the hexagonal MX phases show a negative formation energy (calculated with respect to the elements). Besides InSe, several hexagonal MX compounds, including GaS,[49] GaSe[50] and GaTe,[51] have been synthesized in previous experiments; others are hypothetical for now. We note the presence of $Al_2X_3$ chalcogenides in the literature, with the layered structure of β-$Al_2Te_3$ being just one example,[52] and those phases would be competing when attempting to synthesize any monochalcogenide sample; we still include the hypothetical aluminum species in our computational survey, not least for comparison with the (existent) Ga homologues. According to high-level electronic-structure calculations, using a hybrid DFT functional and including spin–orbit coupling (Methods), all nine alloys considered here are semiconductors with either indirect or direct band gaps (Figures 3b and S3). The band gap $E_g$ values span a wide range from 0.6 to 2.3 eV (Table 1), covering parts of the spectrum of infrared and visible light. The calculated $E_g$ of β-InSe is 0.96 eV, in fair agreement with the experimental value of 1.2 eV.[30]

To understand the nature of the chemical bonding in these phases, and how this bonding may be tuned, we carried out systematic analyses using the crystal orbital Hamilton populations (COHP) method.[53] Starting from a self-consistent electronic-structure calculation, COHP analysis separates orbital-pair-wise interactions into stabilizing bonding contributions (shown as positive −COHP) and destabilizing antibonding ones (negative −COHP). Being defined for pairs of orbitals on (neighboring) atoms, the method allows us to investigate the interaction between a specific pair of atoms in any given structure. As shown in Figure 3 and Figure S4, the short intralayer homopolar M−M bonds (2.44−2.80 Å) and heteropolar M−X bonds (2.32−2.88 Å) show strong covalent bonding nature, as there is no antibonding interaction at the Fermi level $E_F$, and bonding interactions dominate for the occupied bands in all the MX alloys. Regarding *interlayer* interactions, there are only minor COHP interactions between interlayer M and X atoms, and largely a cancellation of filled bonding and antibonding levels—all being consistent with the large interatomic distances of 4.16−4.89 Å and the presumably van der Waals nature of the interlayer interactions (Figure 3b and Figure S4). The absence of antibonding contributions for both intra- and interlayer interactions near the Fermi level suggests that all the MX phases considered here are stable from a chemical-bonding perspective. Moreover, the magnitudes of −COHP for intralayer M−X and M−M bonds are much larger than those of interlayer M−X interactions, indicating that intralayer bonds are much stronger than the interlayer interactions, which can maintain the intralayer structures along with the interlayer slip.

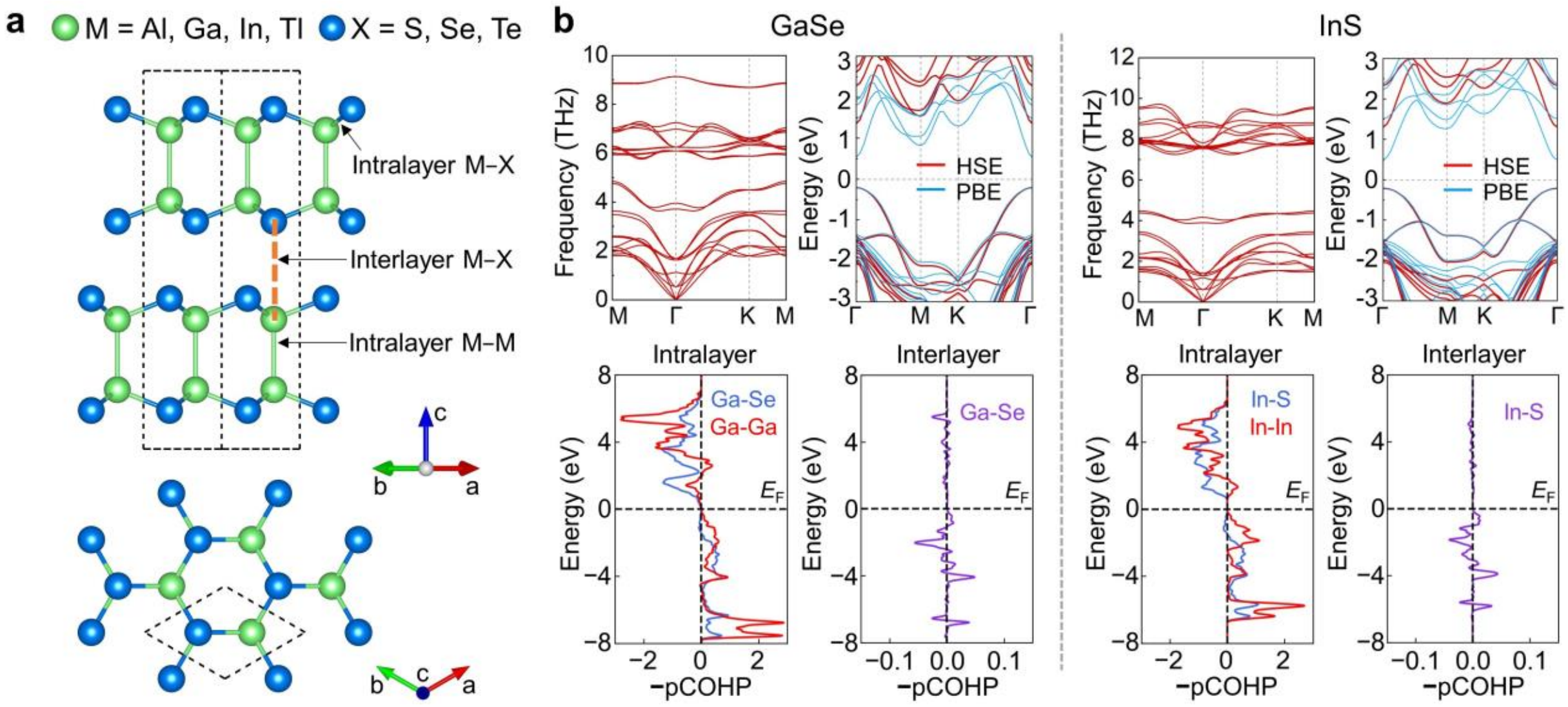

**Figure 3. Density functional theory calculations for group-III chalcogenides MX. a** Atomic structure for MX (M=Al, Ga, In; X=S, Se, Te). Green and blue spheres represent M and X atoms, respectively. The gold dashed line across the interlayer gap highlights the minimum interlayer interaction between the neighboring layers, which is the interlayer M−X interaction. **b–c** Calculated phonon dispersion, electronic band structure and projected COHP analysis for GaSe and InS, respectively.

**Table 1.** Computed properties of known and hypothetical group-III chalcogenides in the same structure as β-InSe.

| | Lattice parameters (Å) | | Intralayer M–X/M–M distance (Å) | Interlayer M–X distance (Å) | Layer distance (Å) | Band gap (HSE06[#]) (eV) |
|---|---|---|---|---|---|---|
| | *a* = *b* | c | | | | |
| AlS | 3.56 | 15.94 | 2.32 / 2.58 | 4.33 | 3.27 | 2.16, indirect |
| AlSe | 3.76 | 16.50 | 2.46 / 2.56 | 4.53 | 3.37 | 1.79, indirect |
| AlTe | 4.10 | 17.42 | 2.69 / 2.55 | 4.89 | 3.62 | 1.39, indirect |
| GaS* | 3.62 | 15.61 | 2.36 / 2.46 | 4.26 | 3.17 | 2.31, indirect |
| GaSe* | 3.80 | 16.12 | 2.49 / 2.45 | 4.44 | 3.27 | 1.57, direct |
| GaTe* | 4.12 | 16.98 | 2.69 / 2.44 | 4.78 | 3.52 | 0.60, indirect |
| InS* | 3.92 | 16.28 | 2.55 / 2.80 | 4.16 | 2.98 | 1.52, direct |
| InSe* | 4.07 | 16.85 | 2.68 / 2.79 | 4.36 | 3.08 | 0.96, direct |
| InTe | 4.36 | 17.76 | 2.88 / 2.78 | 4.72 | 3.34 | 0.68, direct |

* These materials have been experimentally synthesized in the same structure as β-InSe. (In contrast, the Al phases are hypothetical, and InTe is known although in a different structure.)

[#] Spin–orbit coupling has been considered (Methods).

Intuitively, what we desire is a synergy between a sufficiently low slipping energy $E_s$ and an adequately high cleavage energy $E_c$, both of which can be calculated by adopting the method proposed in Ref. [16] for all the nine MX alloys. As shown in Figure S5, we divided each slipping period into 12 steps, and calculated the energy variation with respect to the interlayer distance for each slipping step. $E_s$ and $E_c$ are defined as the maximum energy difference (i.e., maximum energy required) along the slipping, and the minimum energy difference (i.e., maximum energy required) to fully exfoliate the layer, respectively. All the calculated $E_s$ and $E_c$ are shown in **Figure 4**a. The $E_c$ values range from 12.8 (AlTe) to 14.8 meV/Å$^2$ (GaSe), and the $E_s$ values are from 1.6 (AlS) to 2.7 meV/Å$^2$ (InS). Although InSe possesses nearly the largest $E_c$, which is beneficial for structural integrity against intralayer debonding/disintegration, its $E_s$ is among the largest ones as well, which would not be conducive to interlayer sliding deformation. Importantly, as the cation-like species changes from In to Ga and Al, the $E_c$ values are only reduced by ~10%, while the $E_s$ values can be reduced by as much as ~40%, implying that the constituent elements have a larger effect on $E_s$ than on $E_c$. We then assessed the correlation between these two parameters and the interlayer interactions. Given the same structure type for all the homologues, the degree of covalent bonding can be evaluated by the integrated −COHP (−ICOHP) along the energy axis up to $E_F$, which is taken to be a measure for the bond strength. Therefore, we mapped $E_c$ and $E_s$ with the −ICOHP of interlayer interactions (Figure 4a), which shows that both $E_c$ and $E_s$ depend on the −ICOHP values, i.e., a stronger interlayer interaction would resist slip deformation while promoting structural integrity.

Similarly, we also calculated the in-plane Young's modulus $E_{in}$ and the fracture toughness $K$, and related the two parameters with the −ICOHP of intralayer bonding (Figure 4b). The $E_{in}$ values (42−100 GPa) and $K$ values (0.14−0.32 MPa·m$^{0.5}$) show a strong positive correlation. Alloys of lighter elements, such as InS or GaSe, are more resistant to fracture, whilst sacrificing elastic features to some extent. The calculated $K$ values of all nine alloys are larger than those of typical layered thermoelectric compounds such as $Bi_2Te_3$ and SnSe (< 0.03 MPa·m$^{0.5}$).[48] As for a bonding–property relationship, both $E_{in}$ and $K$ are correlated with the −ICOHP values of intralayer M−X bonds, since stronger bonds would stiffen the compounds and enlarge the energy cost for bond breaking. Note that the intralayer M−M interactions have only marginal influence on $E_{in}$ and $K$, since these directional bonds are perpendicular to the stress directions.

Based on the above analysis, we now move on to identify the likely best performing material for practical forming applications. We first evaluated the deformability for the MX compounds considered using the "deformability factor" $\Xi = \frac{E_c}{E_s} \cdot \frac{1}{E_{in}}$ proposed in ref. [16]. As shown in Figure 4c, most of the MX alloys are predicted to possess a deformability comparable to that of InSe. The $\Xi$ values for all the MX are consistently larger than that of graphene and $MoS_2$, whose $\Xi$ values are only ~20% of that of InSe.[16] However, this $\Xi$ parameter is not sufficiently discriminative to tell apart the fracture tendency of these candidate compounds. As we have discussed earlier, the fracture toughness is another critical property that determines the feasibility of rolling processing. We therefore constructed a $K$–$\Xi$ map (Figure 4d). This map explicitly reveals that there is a large disparity in fracture toughness: the Al- and Ga-based chalcogenides possess much larger fracture toughness than InSe, while exhibiting a $\Xi$ comparable to InSe. This observation strongly hints at a possibility to improve the forming ability when InSe is substitutionally alloyed with Ga or (where this is feasible) with Al.

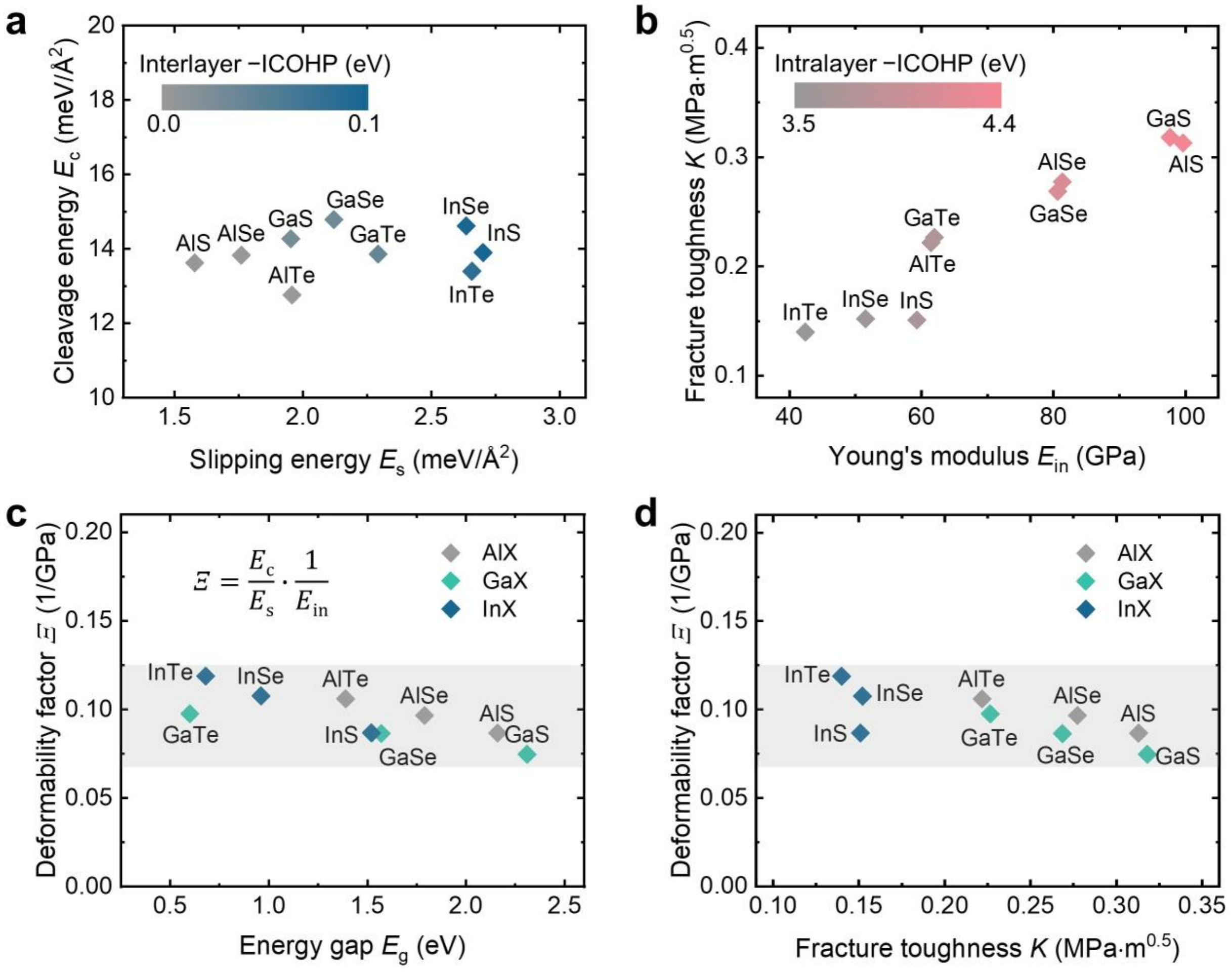

**Figure 4. Property predictions for MX alloys.** **a** Calculated $E_s$ and $E_c$ values for MX in terms of −ICOHP values of interlayer M−X bonding. **b** Calculated in-plane Young's modulus $E_{in}$ and fracture toughness values $K$ for MX in terms of −ICOHP values of intralayer M−X bonding. The colorbar in **a** and **b** indicates the magnitude of the −ICOHP values, which is taken as a measure for the covalent bond strength. **c** Deformability factor $\Xi$ and band gap $E_g$ for MX. The deformability factor $\Xi$ is defined as $\Xi = \frac{E_c}{E_s} \cdot \frac{1}{E_{in}}$. **d** Deformability factor $\Xi$ plotted against fracture toughness values $K$ for MX alloys.

To validate our prediction, we synthesized single-crystalline GaSe samples using the same method as that for InSe (see Methods section). The GaSe sample is initially 4 mm in length and 0.410 mm in thickness (**Figure 5**). We followed the same roller pressing route as that used for InSe. The GaSe sample can be readily thinned without failure from 0.410 mm to 0.067 mm in thickness, showing excellent rolling plasticity. A recent work also shows the large plasticity of submicrometer-scale single-crystalline GaSe.[54] Typical SEM images of the surfaces after rolling are shown in Figure 5a (InSe) and Figure 5b (GaSe). The yellow arrows highlight the cracks, while the white arrows mark the edges of the slipped layers that leave terraces. The major difference between the cracks and layer edges is that the edges are commonly very straight and long (across the width of the sample), while cracks are curvy and short. Since the cracks are always accompanied by narrow slits/pits, we define the slits whose widths are wider than 100 nm as already-initiated cracks. Sampling the number of cracks on various random positions of the two rolled samples (Figure 5 and Figure S6), we found that there were 11.2 cracks per 100 μm$^2$ for InSe, while there were only 0.8 for GaSe. Such an obvious contrast in the crack density unambiguously validates our prediction that GaSe has larger fracture resistance than InSe, and meanwhile its slip deformability is not degraded with respect to that of InSe.

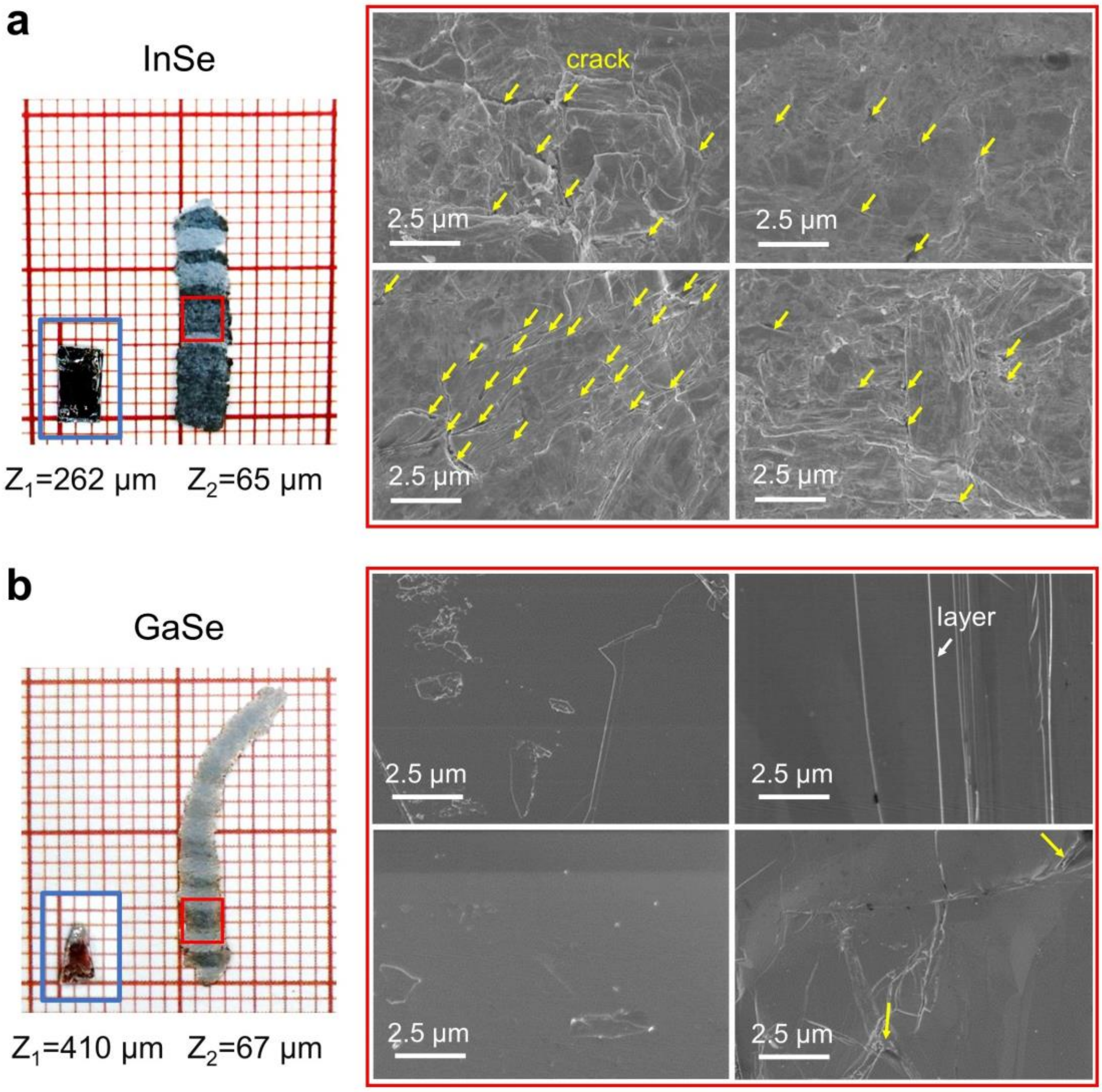

**Figure 5. Morphology of rolled samples. a–b** The InSe and GaSe samples, respectively. The left panel shows the sample before and after rolling (the inset blue box shows the samples before rolling; the grids are 1 × 1 mm$^2$ in size). $Z_1$ and $Z_2$ indicate the thickness of the sample before and after rolling. The right panel shows the SEM images of the red-boxed regions in the left panel. The white and yellow arrows mark the edge of layers and the initiated cracks, respectively.

Before closing, we present two materials maps to comment on potential changes in electronic and thermoelectric properties, as several materials in the MX family, such as InSe, GaS, GaSe and GaTe, are promising candidates for practical applications.[27-30, 55-57] **Figure 6**a shows energy gap values $E_g$ in terms of $\Xi$ and $K$. The $E_g$ values decrease with $\Xi$, and increase with $K$. As for the thermoelectric application, $zT$ values are regarded as the figure of merit to represent the efficiency of the device. In ref.[58], all the $zT$ values for monolayer MX have been calculated based on DFT and Landauer approach. According to Hicks–Dresselhaus theory,[59] the values for the two-dimensional form can be treated as the upper limit for the bulk counterparts. Here, we used the monolayer $zT$ values to estimate the tendency of $zT$ for bulk MX. As shown in Figure 6b, the $zT$ value increases with $\Xi$, and decreases with $K$. For example, we can simultaneously improve the thermoelectric property and deformability by sacrificing a bit fracture toughness by alloying InSe with Te. It has been demonstrated that alloying InSe with Te can indeed enhance the $zT$ value.[38] We can also alloy InSe with Ga to increase the fracture toughness, while sacrificing $zT$ value to some extent, if the goal is to improve the forming ability. We note that these maps are based on ideal single crystalline phase. Certain extra effects of alloying may occur beyond these maps. For instance, the point defect scattering and high-entropy effect induced by alloying could result in extra enhancement of thermoelectric properties.[60-62] Alloying may also induce a secondary phase[63] or asymmetrical structure[64] which can effectively impede crack propagation and enhance fracture toughness.

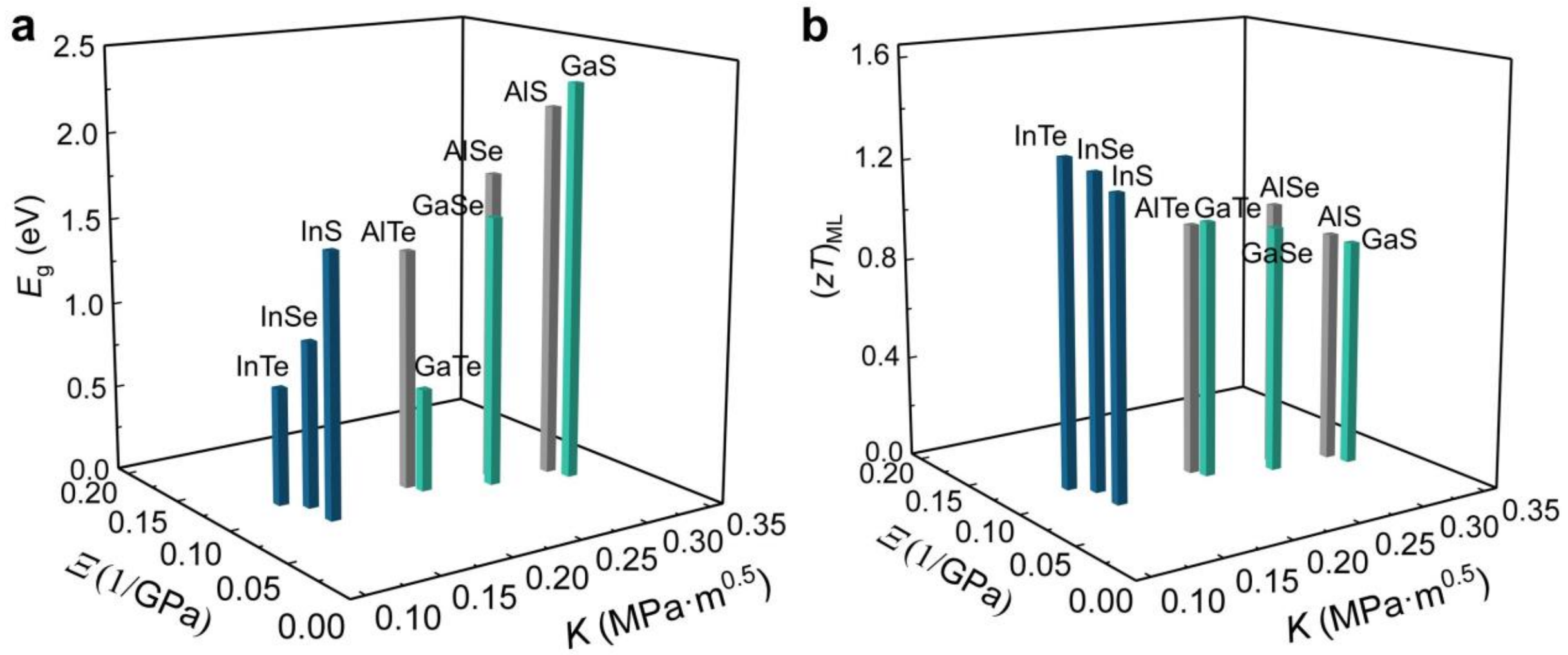

**Figure 6. Functional properties of the chemically-designed MX alloys. a** Energy gap values $E_g$ plotted together with $\Xi$ and $K$. **b** Thermoelectric figure of merit $zT$ values for monolayer MX plotted together with $\Xi$ and $K$.

In summary, our experiments indicate that crystalline InSe, while plastically deformable to some extent, suffers from inadequate resistance against fracture when subjected to rolling. Therefore, to impart forming capability to bulk inorganic semiconductors, we advocate that extra attention should be paid to intralayer fracture toughness (energy required before the onset of damage), via designing appropriate alloying elements to tune the chemical bonds. Based on our model calculations of group III chalcogenides, all the InSe homologues have comparable deformability via interlayer slip. Chemical bonding analyses unveil that the intralayer heteropolar bonds are important in determining fracture toughness, i.e., stronger intralayer bonds would enhance the fracture toughness by resisting bond breaking. We provide materials maps for simultaneously reaching the desired deformability and fracture toughness, identifying GaSe as an example of a rolling-friendly van der Waals semiconductor. Overall, a multi-parameter alloy design strategy is required, in dealing with the trade-off among the electrical, thermoelectric and mechanical properties. As a start in this direction, our work above serves as a guide for screening and designing intrinsically ductile inorganic semiconductors.

## Methods

*Materials and characterization***.** Single-crystalline InSe and GaSe samples were prepared using the Bridgman method (commercially available from the Nanjing MKNANO Tech. Co., Ltd.). The crystal structures of the as-grown InSe and GaSe samples were characterized by X-ray diffraction (Bruker D8 Advance, Bruker AXS) with Cu $K_{\alpha}$ source, and it confirmed that the two samples are both in the same structure type as β-InSe. The InSe and GaSe single crystalline sheets were cold rolled at room temperature with the compression direction along the *c*-axis of the crystal. The thickness of the InSe sample was reduced from 0.262 to 0.065 mm after the rolling compression, and the GaSe sample was thinned from 0.410 to 0.067 mm. The surface morphology of InSe and GaSe crystals before and after rolling was characterized using a Hitachi SU8230 SEM. We mechanically cleaved intact bulk InSe crystals with scotch tape to obtain flakes with clean surfaces and crack edges. The flakes were then transferred onto a PDMS-based gel (Gel-Pak® PF film). We located the flakes under an optical microscope. Then we transferred the flakes onto a TEM grid (SiN grid) for further characterization in a JEOL JEM-2100F TEM operating at 200 kV. Both high resolution TEM (HRTEM) imaging and fast Fourier transform (FFT) of HRTEM images were used to distinguish zigzag- and armchair-type edges.

*Computational methods*. Density functional theory (DFT) calculations were carried out using the Vienna Ab Initio Simulation Package (VASP)[65] with the projector augmented-wave (PAW) method[66] within the generalized gradient approximation (GGA).[67] The D3 scheme was used to include van der Waals corrections.[68] All structures were fully relaxed with regard to both atoms and lattice parameters. The relaxed lattice parameters for β-InSe are 4.07 and 16.85 Å for *a* (=*b*) and *c*, respectively, which matches well with the experimental values (*a*=*b*=4.01 Å, *c*=16.67 Å).[69] HSE06 hybrid functional calculations[70] combined with spin–orbit coupling (SOC) were used to calculate the electronic band structures. The cutoff energy was set to 700 eV and the *k*-point mesh was 16 × 16 × 4. Phonon dispersions were calculated by the finite displacement method as implemented in Phonopy package.[71] A 4 × 4 × 1 supercell expansion with a *k*-point mesh of 4 × 4 × 4 was used to calculate interatomic force constants. Crystal orbital Hamilton population (COHP) analyses were carried out using the LOBSTER code.[72-74] The surface formation energy is defined as $\Delta E = (E_{slab}-E_{bulk})/2S$, where $E_{slab}$ and $E_{bulk}$ are the total energies of the slab and bulk models, respectively and $S$ is the area of the fractured surface of the slab model. The ideal stress–strain curves for biaxial tensile loading were calculated via DFT, as developed in ref. [46-48]. The tensile deformation was imposed along two directions, while the other strain components could fully relax (both lattice and atom positions). The stress $\sigma$ was obtained according to $\sigma = \frac{1}{V_0}\frac{\partial E}{\partial \varepsilon}$, where $V_0$ is the initial volume, $E$ is the static energy and $\varepsilon$ is the tensile strain. To obtain the ideal stress–strain curves, we set the interval of strain increment to 1%, and each static energy calculation of incremented strain was based on the relaxed structure of the previous step.